\documentclass[conference]{IEEEtran}
\IEEEoverridecommandlockouts
\usepackage{cite}
\usepackage{amsmath,amssymb,amsfonts}
\usepackage{algorithmic}
\usepackage{graphicx}
\usepackage{textcomp}
\usepackage{xcolor}
\usepackage{hyperref}
\def\BibTeX{{\rm B\kern-.05em{\sc i\kern-.025em b}\kern-.08em
    T\kern-.1667em\lower.7ex\hbox{E}\kern-.125emX}}

\usepackage{pifont}
\usepackage[colorinlistoftodos,textwidth=1.22cm]{todonotes}
\usepackage{xargs}
\usepackage{soul}
\usepackage{wasysym}
\usepackage{float}

\newcommand{\cmark}{\ding{51}}%
\newcommand{\xmark}{\ding{55}}%
\newcommand{\tool}{\texttt{THREAT/crawl}}

\newcommandx{\latodo}[2][1=]{\todo[linecolor=purple,backgroundcolor=purple!25,bordercolor=purple,#1]{#2}}
\newcommandx{\mctodo}[2][1=]{\todo[linecolor=teal,backgroundcolor=teal!25,bordercolor=teal,#1]{#2}}
\newcommand{\req}[1]{\texttt{R{#1}}}
\newcommand{\arch}[1]{\texttt{A{#1}}}

\newcommand{\xss}{\texttt{xss}}
\newcommand{\nulled}{\texttt{nulled}}
\newcommand{\crdclub}{\texttt{crdclub}}
\newcommand{\darknetcity}{\texttt{darknetcity}}
\newcommand{\altenen}{\texttt{altenen}}
\newcommand{\deeptor}{\texttt{deeptor}}
\newcommand{\nulledbb}{\texttt{nulledbb}}

\usepackage{siunitx,tabularx,ragged2e,booktabs,caption}
\usepackage{graphicx,tikz}
\newcommand*\annotatedFigureBoxCustom[8]{\draw[#5,line width=2pt,rounded corners] (#1) rectangle (#2);\node at (#4) [fill=#6,thick,shape=circle,draw=#7,inner sep=2pt,font=\sffamily,text=#8] {\textbf{#3}};}
\newcommand*\annotatedFigureBox[4]{\annotatedFigureBoxCustom{#1}{#2}{#3}{#4}{red}{black}{white}{white}}

\newenvironment {annotatedFigure}[1]{\centering\begin{tikzpicture}
\node[anchor=south west,inner sep=0] (image) at (0,0) { #1};\begin{scope}[x={(image.south east)},y={(image.north west)}]}{\end{scope}\end{tikzpicture}}

\newcommand*\encircle[1]{\tikz[baseline=(char.base)]{
            \node[thick,shape=circle,fill,inner sep=1pt,font=\sffamily] (char) {\textcolor{white}{#1}};}}
            
\usepackage{array} 
\usepackage{multirow}

\newcounter{hypocounter}
\renewcommand\thehypocounter{(R\arabic{hypocounter})}
\newenvironment{hypothesis}{
\refstepcounter{hypocounter}
\begin{center}
\begin{tabular}{m{0.75\linewidth}c} 
}{&\thehypocounter\end{tabular}\end{center}}

\begin{document}

\title{\tool: a Trainable, Highly-Reusable, and Extensible Automated Method and Tool to Crawl Criminal Underground Forums}

\author{\IEEEauthorblockN{Michele Campobasso}
\IEEEauthorblockA{m.campobasso@tue.nl\\ 
Eindhoven University of Technology\\
Eindhoven, The Netherlands\\
}
\and
\IEEEauthorblockN{Luca Allodi}
\IEEEauthorblockA{l.allodi@tue.nl\\
Eindhoven University of Technology\\
Eindhoven, The Netherlands\\
}}

\maketitle

\begin{abstract}
Collecting data on underground criminal communities is highly valuable both for security research and security operations. Unfortunately these communities live within a constellation of diverse online forums that are difficult to infiltrate, may adopt crawling monitoring countermeasures, and require the development of ad-hoc scrapers for each different community, making the endeavour increasingly technically challenging, and potentially expensive. To address this problem we propose \tool, a method and prototype tool for a highly reusable crawler that can learn a wide range of (arbitrary) forum structures, can remain under-the-radar during the crawling activity and can be extended and configured at the user will. We showcase \tool\ capabilities and provide prime evaluation of our prototype against a range of active, live, underground communities.
\end{abstract}


\section{Introduction}


Underground cybercrime communities are increasingly more important to understanding and measuring the overall threat landscape. Security operators or security service providers scrape them to obtain key intelligence on emerging threats~\cite{bouwman2020different}; law enforcement scrape (and sometimes run) them to monitor cybercrime operations and networks~\cite{europol}; security researchers are interested, among other things, in understanding the dynamics of attack innovation~\cite{allodi2017economic,anderson2021silicon}, identify key actors operating in these forums~\cite{pastrana2018characterizing}, or investigate novel or emergent threats~\cite{hutchings2019understanding, campobasso2020impersonation}.
The landscape of underground criminal forums is large, with dozens of forums operating in the English, Russian, and Chinese spheres -- among many others~\cite{pastrana2018crimebb,allodi2017economic,desombre2021primer}; new communities appear continuously, existent ones evolve, and the more prominent communities deploy anti-crawling technologies to block scraping activities, and ban (sometimes hard to obtain) related access credentials~\cite{pastrana2018crimebb,campobasso2020impersonation,campobasso2019caronte}.
This makes the endeavour of monitoring cybercrime communities time consuming, technically challenging, and expensive to run. A large fraction of this overall ``cost'' is constituted by the need to build \textit{ad-hoc} crawlers and parsers capable of correctly navigating different forums, extract relevant content, while remaining under the radar~\cite{pastrana2018crimebb}.

In this paper we first derive from the literature and discuss the overall ``foundational'' dimensions of this problem, and identify the requirements for a general method over which we design our prototype, \tool. We showcase several design choices that can effectively bridge the gap between the dimensions of the problem, allowing to develop a tool that can learn how to crawl a wide range of different forum structures without requiring its users to re-write a parser for each different forum (or, sometimes, forum section). \tool\ provides a simple interface for users to identify specific elements of interest, such as navigation buttons and content, and employs a set of strategies to automatically instrument the crawling engine with the corresponding information needed to traverse the HTML structure of the relevant page(s). Further, \tool\ can be extended by allowing users to inject (JS) code during the training/crawling phase to perform specific actions (e.g., to adapt the procedure to the specific forum instance), and comes by design with extensive stealth capabilities to remain under the radar during crawling, if needed. To evaluate the effectiveness of our overall method and of its prototype implementation, we showcase the implemented features of the tool against a set of live, active underground communities and discuss functionalities and limitations of our implementation. \tool\ will be released publicly and freely upon publication of this paper.

The contribution of this paper is threefold:
\begin{itemize}
    \item We analyze the foundational challenges posed by the problem of designing a general, reusable crawler for underground forums; the identified challenges can help frame future contributions in this space;
    \item We propose a general method and solution to address the identified challenges, and provide an implementation showcasing the method and the architectural components addressing each of the identified challenges;
    \item We evaluate our method and prototype tool against seven live, active criminal underground forums and identify key aspects for improvement of the tool implementation. We release \tool\ publicly to allow for any uptake and employment of the tool from the community.
\end{itemize}

This paper proceeds as follows: in Section~\ref{sec:probspace} we identify key dimensions of the problem and derive requirements that a general solution must satisfy; Section~\ref{sec:relwork} discusses related work and compares different solutions over the identified requirements. Section~\ref{sec:solution} details the overall design and implementation of our solution, and Section~\ref{sec:evaluation} evaluates it against seven live underground communities. Section~\ref{sec:discuss} discusses results and limitations of our solution, the next steps to take to evolve \tool\ to a mature solution, and Section~\ref{sec:conclusions} concludes the paper.

\section{Problem space and solution requirements}
\label{sec:probspace}
We identify two main dimensions to the problem of designing a general, reusable crawler for underground forums: the diversity of the forums, and the adversarial nature of the monitored environment. A third dimension on ethical aspects is transversal to these.

\subsection{Adversarial environment for crawling}
\label{subsec:prob_netmon}
Crawlers can oftentimes be easily identified due to their high content fetch rates from a target, and from their typical approach to explore the available content of the target website.
These include consumer services such as Cloudflare DDoS protection~\cite{turk2020tight, decary2015sifting} and other DDoS protection services, provided from underground actors or by so-called Bulletproof-Hosting services~\cite{aditya2013crimeware, goncharov2012russian}, specialized in defending onion websites. 
Browser fingerprinting is an effective strategy requiring the request issuer to execute JavaScript to verify a number of properties of the browser environment, which are hard to mimic with the use of scripts. 
Other anti-crawler measures include HTTP requests inspection, which can provide several indicators of bot activity (e.g., lack of proper ``referer'' headers in HTTP requests).
As a result of the detection, the forum may slow down the crawling process~\cite{turk2020tight}, showing CAPTCHAs~\cite{decary2015sifting} or throttling traffic, or may ban the forum account used to access the crawled resources~\cite{turk2020tight}. The latter case is particularly concerning in the case of communities enforcing strict access control mechanisms at registration time (e.g., registration on invitation or paywall), potentially jeopardizing months or years of efforts in creating a ``legitimate'' identity in the underground to infiltrate~\cite{campobasso2020impersonation}; similarly, this may pose ethical issues when access fees have to be paid multiple times, potentially compromising the balance between achieving research goals and not providing tangible (economic) support to criminals as a result of multiple payments. 
Hence, to deal with anti-crawler countermeasures, a general tool has to be:
\setcounter{hypocounter}{0}
\begin{hypothesis}
\textbf{Stealth}: it should avoid generating suspicious traffic, while attempting to reproduce the forum navigation activity of a regular user.
\end{hypothesis}

\noindent
Depending on the community to monitor, researchers may want to finely-tune the time at which pages are visited (e.g. according to specific time-zones) and more in general the crawling operation as a whole by gathering only relevant information from specific sections.
Therefore, a general tool must also be:
\begin{hypothesis}
\textbf{Configurable}: it should allow the user to finely tune the crawler operation, including the speed and time of the crawling and providing additional (run-time) information for its execution.
\end{hypothesis}


\subsection{Diverse underground communities}
\label{subsec:prob_diverse}
Albeit most forums are generally similar to each other, implementations come with their own peculiarities that make (automated) navigation not trivial~\cite{pastrana2018crimebb, turk2020tight, fallmann2010covertly, fang2019analyzing, nunes2016darknet, decary2015sifting, benjamin2019dice}. To scrape their pages, developing an \textit{ad-hoc} crawler for each target website is a costly and inefficient procedure, which has to account for different aspects of the target~\cite{fang2019analyzing, jiang2012focus, turk2020tight}. Forums may be implemented using multiple CMSes~\cite{fallmann2010covertly, jiang2012focus, turk2020tight}, with different versions, flavors and skins, or even be completely custom solutions~\cite{pastrana2018crimebb, turk2020tight}, making the derivation of a crawling algorithm addressing all the peculiarities for each forum a challenging task. Also, several forums implement anti-crawler mechanisms that randomize HTML attributes like IDs or classes, making the position of content within a page unpredictable. Hence, a general crawler tool should be:
\setcounter{hypocounter}{2}
\begin{hypothesis}
\label{req:train}
\textbf{Trainable}: it should be capable of learning how to crawl different forums,
independently of their structure, DOM properties, and design and deployment solutions.
\end{hypothesis}

\smallskip
\noindent Because of non-standard implementations that may appear across different forums (e.g., customization of navigation features such as JavaScript-enabled navigation buttons), a general tool must also be:
\begin{hypothesis}
\label{req:extend}
\textbf{Extensible}: it should allow users to extend the tool capabilities by injecting simple procedures in the crawling process whenever a non-standard situation not supported by the tool is encountered.
\end{hypothesis}
\smallskip
Apart from the challenge of crawling different forums, merely downloading pages still poses the problem of content extraction~\cite{decary2015sifting}, as it may be organized differently across different forums. For example, each post in a thread generally contains information regarding the author (e.g., registration date, popularity, number of posts, ...), but their arrangement and/or identification in the page may vary significantly, some may be missing, and their position may not be constant across sections of the same forum. This forces the implementation of custom parsers for content extraction for each single forum to scrape (and, occasionally, different parsers for pages in the same forum)~\cite{turk2020tight, nunes2016darknet}. This adds additional overhead to the data collection process, increases time-consuming testing requirements, and generates parser software that cannot generally be re-used. Therefore, a general tool should offer: 
\begin{hypothesis}
\label{req:parsing}
\textbf{Structured data collection}: it should provide the capability to parse content from crawled pages regardless of how these are structured.
\end{hypothesis}

\subsection{Ethical considerations}
Usage of a crawler should always be subject to ethical considerations. Given the sensible nature of the problem addressed in this paper (i.e., monitoring criminal forum communities), we believe these concerns should be addressed at the design level by the tool itself rather than being left entirely for further consideration of the user. As such, requirements over this dimension are to an extent `orthogonal' to the other requirements, and we therefore label them differently as meta-requirements, \texttt{MRx}.

We identify two major concerns that must be addressed. First, the monitoring of adversarial environments may require the user to remain anonymous during (and oftentimes after~\cite{campobasso2020impersonation}) the crawling activity, which in practice often results in tunneling the traffic over the TOR virtual network~\cite{campobasso2019caronte, pastrana2018crimebb, benjamin2019dice, nunes2016darknet, campobasso2020impersonation}. Because of the limited bandwidth available to each Onion Router (i.e. a hop in the TOR network), this may compromise or altogether inhibit the experience of other TOR users (which may be using it to communicate sensible data, or avoid surveillance on their Internet activity). 
Therefore, a general tool should satisfy the following meta-requirement:
\renewcommand\thehypocounter{(MR\arabic{hypocounter})}
\setcounter{hypocounter}{0}
\begin{hypothesis}
\label{metareq:parsimony}
\textbf{Parsimonious}: a tool should limit the bandwidth usage over a private network.
\end{hypothesis}

\noindent Secondly, some content on criminal forums and communities in general may be offensive, or outright illegal even only to access. It is therefore important that a general tool respects the following meta-requirement:
\begin{hypothesis}
\label{metareq:censoring}
\textbf{Censoring}: a tool should be able to censor by not obtaining, download, and save, material that is undesired by the user.
\end{hypothesis}

\noindent
In addition, collecting data from underground markets and using techniques to remain under the radar may violate the terms of services of a target forum~\cite{pastrana2018crimebb} and potentially disrupt their activity~\cite{christin2013traveling}. 
Generally, the societal benefit of studying cybercrime outweighs the remaining risks~\cite{martin2016ethics, benjamin2019dice}, as also evidenced by the numerous studies in this domain, but ultimately this evaluation has to remain with the final user (and relevant ERB).

\section{Related work}
\label{sec:relwork}

Table~\ref{tab:literature}
\begin{table}[t]
\centering
\small
\caption{Mapping of tools and solutions from the literature to the requirements of the problem space.}
\label{tab:literature}
\begin{tabular}{ccccccccc}
\toprule
     & \req{1} & \req{2} & \req{3} & \req{4} & \req{5} & \texttt{MR1} & \texttt{MR2} & Released? \\
\midrule
 \cite{fallmann2010covertly} & \RIGHTcircle & \RIGHTcircle & \Circle & \RIGHTcircle & \Circle & \Circle & \Circle & No \\ %
 \cite{jiang2012focus} & \Circle & \Circle & \RIGHTcircle & \Circle & \Circle & \Circle & \Circle & No \\ %
 \cite{pastrana2018crimebb} & \CIRCLE & \RIGHTcircle & \Circle & \CIRCLE & \CIRCLE & \CIRCLE & \CIRCLE & No \\ %
 \cite{fang2019analyzing} & \RIGHTcircle & ? & \Circle & \Circle & \CIRCLE & \Circle & \Circle & No \\ %
 \cite{cheng2021hackerrank} & \RIGHTcircle & ? & \Circle & \Circle & \RIGHTcircle & \Circle & \Circle & No \\ %
 \cite{benjamin2019dice} & \RIGHTcircle & ? & \Circle & \Circle & \CIRCLE & \Circle & \Circle & - \\ %
 \cite{campobasso2019caronte} & \CIRCLE & \Circle & \CIRCLE & \Circle & \CIRCLE & \RIGHTcircle & \Circle & Yes \\ %
 \cite{nunes2016darknet} & \Circle & \Circle & \Circle & \Circle & \RIGHTcircle & \RIGHTcircle & \Circle & Yes$^\dagger$ \\ %
 \cite{kawaguchi2017ai} & \Circle & \Circle & \Circle & \Circle & \RIGHTcircle & \Circle & \Circle & No \\ %
 
$\star$ & \CIRCLE & \CIRCLE & \CIRCLE & \CIRCLE & \CIRCLE & \CIRCLE & \CIRCLE & Yes$^\ddagger$ \\
\bottomrule
\multicolumn{9}{l}{$^\dagger$: under commercial agreement; $^\ddagger$: after publication.}
\end{tabular}
\end{table}
provides an overview of previous work on (or employing) crawlers for underground forums, alongside some open-source implementations available. 
Early studies collecting data from live underground communities can be traced back to the late 2000s~\cite{zhuge2009studying, fallmann2010covertly}; almost immediately, two problems emerged: the development of complex crawling infrastructures for that specific purpose, and the need to implement strategies to circumvent the target's (at that time, rudimentary) anti-crawler measures~\cite{fallmann2010covertly}. The overhead of creating non-reusable software for each target platform quickly became an evident problem; Jiang et al. proposed a supervised learning tool to teach the crawler how to predict the URL structure of links in a forum~\cite{jiang2012focus} (\req{3}). However, this solution 
disregards the critical aspects of preserving stealth operations, and there are no clear suggestions of data parsing capabilities. Other studies on cybercriminal activities with the use of crawlers to scrape content from underground communities followed, but remained limited in the scope of the analysis to one or few forums, using \textit{ad-hoc} solutions (\req{4}), and not or only partially accounting for stealthiness~\cite{kawaguchi2017ai, nunes2016darknet, fang2019analyzing, jiang2012focus, cheng2021hackerrank} (\req{1}). Other studies rely on not-clearly defined crawling infrastructures, and focus on the development of different data extraction strategies~\cite{fang2019analyzing, kawaguchi2017ai} or on the identification of relevant products~\cite{nunes2016darknet} and actors~\cite{cheng2021hackerrank} using natural language processing (\req{5}). Campobasso et al.~\cite{campobasso2019caronte} developed a software showcasing a supervised procedure to teach the crawler where to find the needed elements to crawl, parse and save within the pages of a target forum (\req{3}, \req{5}), while accounting for some anti-crawler techniques and trying to remain stealth by modelling human behavior~\cite{campobasso2019caronte, pastrana2018crimebb, benjamin2019dice} (\req{1}). A more remarkable example of a reusable software was proposed from Pastrana et al.~\cite{pastrana2018crimebb}, providing support to new modules (\req{4}) to enable crawling and data extraction (\req{5}). They developed a crawler accounting several aspects to conduct stealth operations such as human behavior modelling (\req{1}), similarly to~\cite{campobasso2019caronte}, including also the possibility to enable or disable specific behaviors (\req{2}). However, the software does not offer a guided procedure to the creation of new modules. In the panorama of open-source commercial solutions, we mention some of the most famous general purpose crawlers, such as \textit{scrapy}~\cite{scrapy}, \textit{Apache Nutch}~\cite{apachenutch}, and \textit{Heritrix}~\cite{heritrix}, albeit none of them is designed to stealthily crawl underground communities. Some of these solutions aim at rapidly extract content from pages~\cite{scrapy, apachenutch} (sometimes at the cost of stumble into rate limiting or ban from the target~\cite{scrapy-ban}) or can be used for archival purposes~\cite{heritrix}, they offer their users a great degree of flexibility~\cite{scrapy} while requiring to code the business logic of the crawler (including any anti-detection strategy), and necessitate of additional libraries to support browser capabilities, like session handling or JavaScript execution~\cite{scrapy, apachenutch, heritrix}. 

\section{Overall Method and Solution Design}
\label{sec:solution}

We name our method and tool \tool. Table~\ref{tab:summaryRx} in the Appendix
provides an overview of issues associated with each (meta) requirement, and the corresponding strategy employed by \tool. The overall \tool\ design is described in the following.

\subsection{Solution architecture}
\label{subsec:arch}
 The different architectural components are mapped to one or more requirements; the mapping is summarized in Table~\ref{tab:req}.

\begin{table}[t]
\centering
\small
\caption{Requirements to arch. components mapping.}
\label{tab:req}
\begin{tabular}{lccccc}
\toprule
     & \req{1} & \req{2} & \req{3} & \req{4} & \req{5}  \\
\midrule
\arch{1} Training module & \cmark & \cmark & \cmark &         & \cmark \\
\arch{2} Javascript injection module&        &        &        & \cmark  &        \\
\arch{3} Scheduler & \cmark & \cmark &        &         &        \\
\arch{4} Crawling module& \cmark & \cmark &        &         & \cmark \\
\arch{5} Privacy pass module &\cmark&\cmark&\\
\bottomrule
\end{tabular}
\end{table}


\paragraph{\arch{1}. Training module}

\begin{figure*}
    \centering
    \includegraphics[width=0.94\textwidth,keepaspectratio]{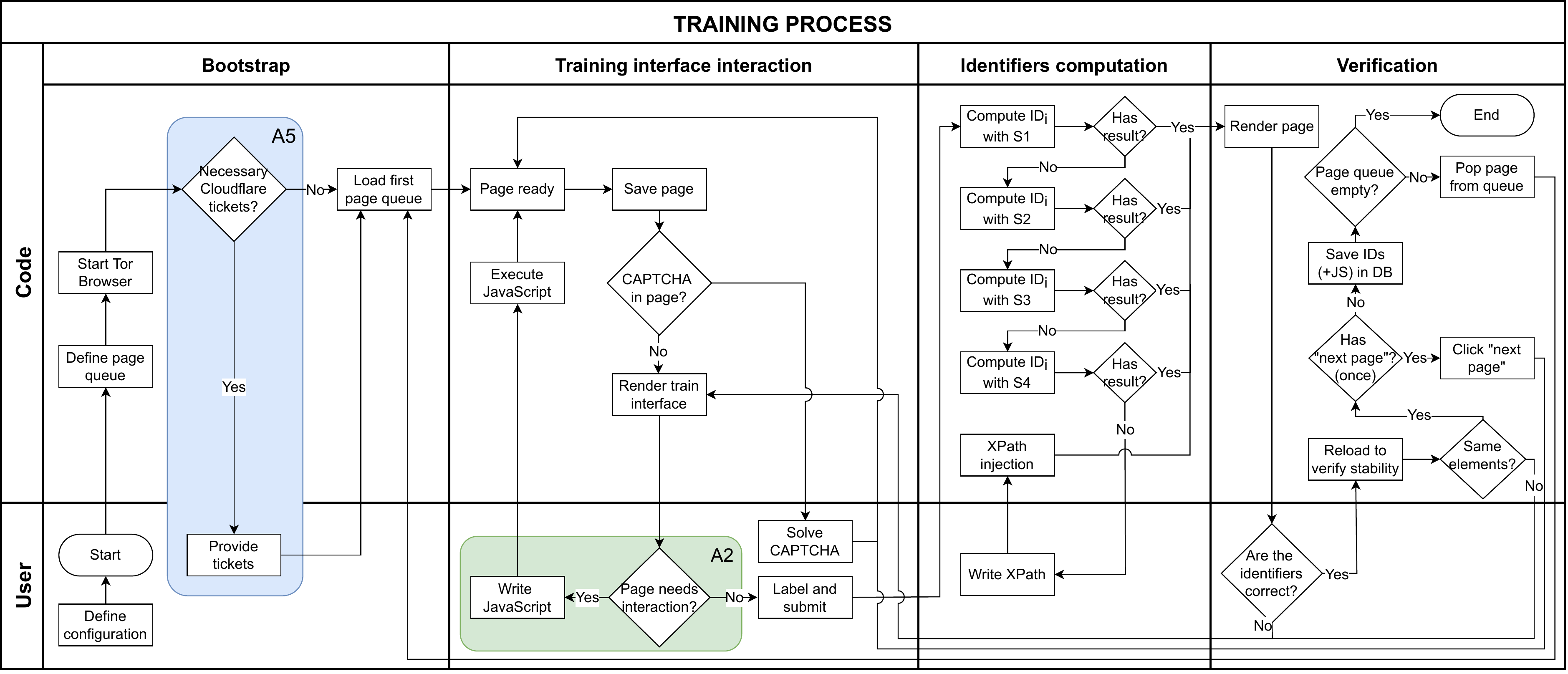}
    \caption{Trainer execution process diagram.}
    \label{fig:train_process}
\end{figure*}

The training process is summarized in the process diagram in Figure~\ref{fig:train_process}. 

\begin{figure}
    \centering
    \includegraphics[width=0.88\columnwidth,keepaspectratio]{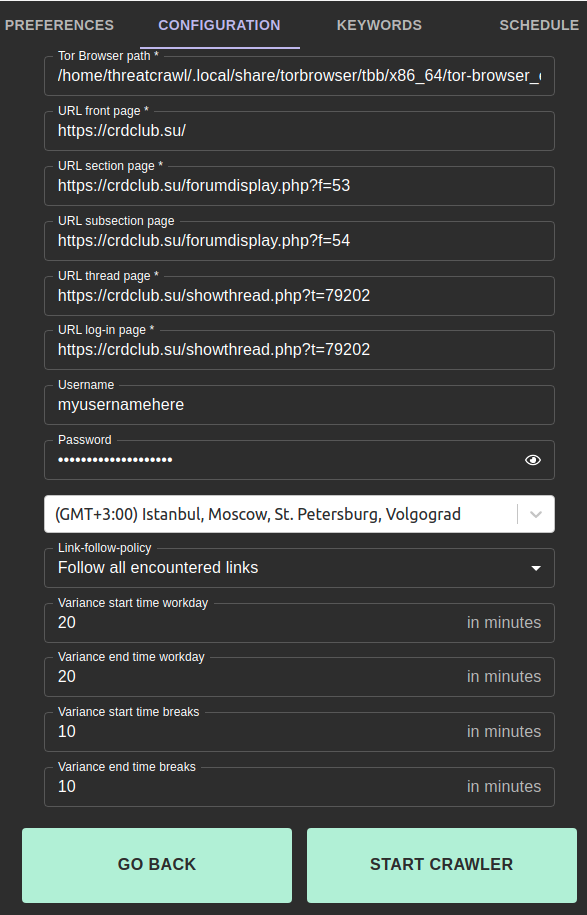}
    \begin{minipage}{0.89\columnwidth}\footnotesize\smallskip
    When creating a new configuration, in the configuration interface the user can provide the list of example URLs that will be used during the training, credentials, timezone to which the schedule will apply, keywords policy (i.e., open only threads containing specific keywords or all), and the start and end variances for both crawler session and breaks in minutes.
    \end{minipage}
    \caption{Configuration interface of \tool.}
    \label{fig:config_interface}
\end{figure}

\smallskip
\noindent
\textit{Bootstrap.} Before crawling a new community for the first time, it is necessary for \tool\ to learn how to navigate it. 
To achieve this, the user starts \tool\ and defines a new configuration, then starts the training. In the configuration interface, the user provides a list of example URLs (login, home page, section, optionally subsection, and thread pages) on which the training will be executed (Figure~\ref{fig:config_interface}). 
In addition, the configuration interface allows to select the desired timezone and variance ranges for the time to start and end of both crawling and breaks. During the configuration it is also possible to specify the policy of content exploration, which can filter out content when matching a list of blacklisted keywords (e.g., avoid exploring threads), or exploring only content containing specific keywords of interest. \tool\ provides an interface to define the blacklisted keywords to filter from the crawling or those of interest (\texttt{MR2}, Figure~\ref{fig:keywords}, in Appendix), and the execution schedule (Figure~\ref{fig:scheduler}, in Appendix). 

\tool\ allows to modify a configuration or a training after these have been defined. The identification of representative example pages is not necessarily an easy task, and changing one of the selected URLs to train again for that page type while preserving the other correct trained pages should be a possible alternative. This is desirable in the case of a minor update in the forum structure breaking the crawling procedure, or when finer tuning is necessary. For this reason, \tool\ offers the possibility to select a previous configuration and untick the ``skip training'' in the configuration interface; starting will result in a new training session pre-labeling the previously identified elements. When started, the tool prepares a queue with the URLs provided during the configuration and starts TOR Browser. The tool asks the user if the target website needs Cloudflare tickets to avoid encountering their DDoS protection page. This will be detailed in the discussion of \arch{5}. The training module can now proceed to load the first page in the queue.

\begin{figure*}
    \centering
    \begin{annotatedFigure}
        {\includegraphics[width=0.97\textwidth,keepaspectratio]{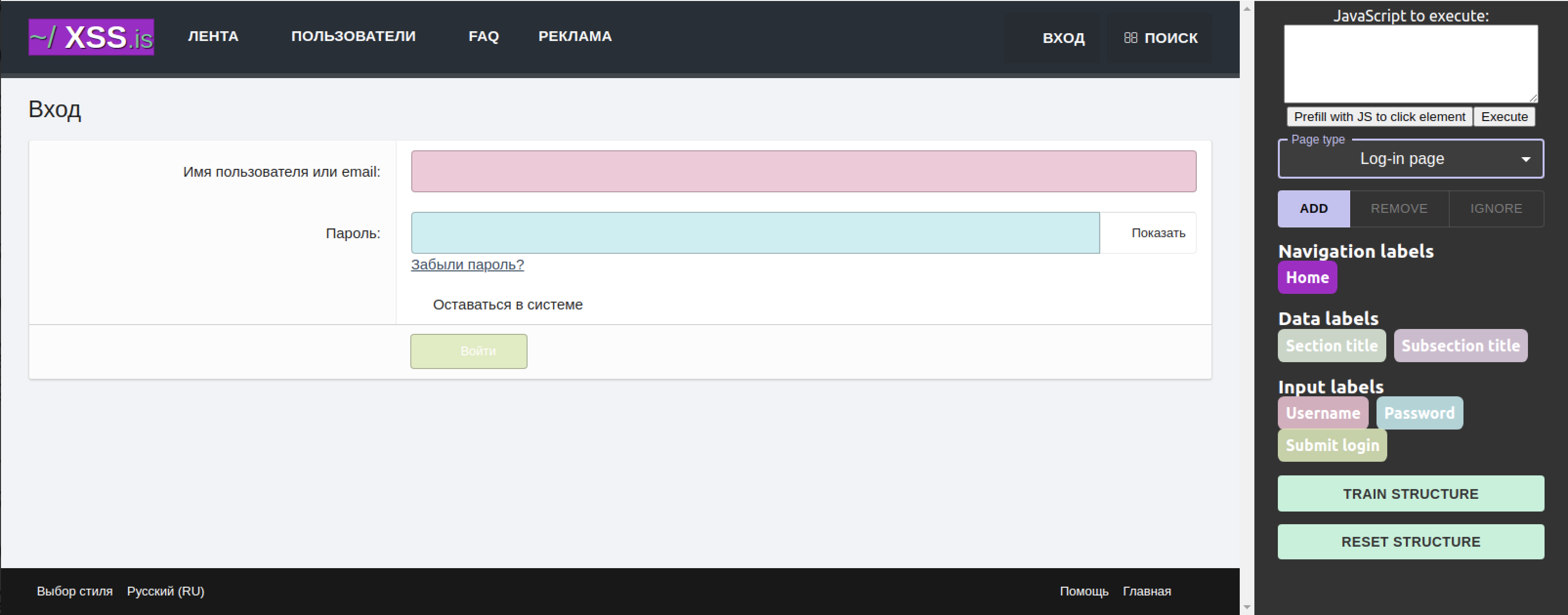}}
	    \annotatedFigureBox{0.002,0.005}{0.7904,0.9948}{A}{0.002,0.0082}
	    \annotatedFigureBox{0.801,0.005}{0.9985,0.9948}{B}{0.801,0.0079}
	    \annotatedFigureBox{0.8201,0.74}{0.82,0.74}{1}{0.8201,0.74}
	    \annotatedFigureBox{0.801,0.6689}{0.80,0.6631}{2}{0.805,0.66}
	    \annotatedFigureBox{0.8,0.3335}{0.8,0.33}{3}{0.805,0.3335}
	    \annotatedFigureBox{0.266,0.72}{0.266,0.72}{4}{0.266,0.72}
    	\annotatedFigureBox{0.839,0.2051}{0.839,0.2016}{5}{0.839,0.2051}

    \end{annotatedFigure}
    \begin{minipage}{0.97\textwidth}\footnotesize
    To the left, there is the rendered page \encircle{A}, and to the right there is the training pane \encircle{B}. From the latter the user first has to specify what type of page is being trained \encircle{1} (in this case, login page). 
    The user selects ``add'' \encircle{2} and clicks on the label to assign (in this case, ``Username'') \encircle{3} in the training pane \encircle{B}. In the rendered page \encircle{A} it is now possible to click on the corresponding input field \encircle{4}. 
    This will color the input field with the same color of the selected label. In the case the user misclicks the box, they can click on ``remove'' next to ``add'' to remove the wrong selection and try again. When the user labeled all the relevant elements, they can confirm by clicking on ``Train structure'' \encircle{5}.
    \end{minipage}
    \caption{Training window for a login page.}
    \label{fig:training_login}
\end{figure*}

\smallskip
\noindent
\textit{Training interface interaction.} When the page is ready in the browser, it is saved and checked for CAPTCHAs. If any, the user is notified via the CLI, asking to solve it. The page is then rendered in the training interface, as shown in Figure~\ref{fig:training_login}. 
The login page is then saved and, if no CAPTCHA was encountered, rendered in the training interface, as shown in Figure~\ref{fig:training_login}.
Before the labeling, the user has the possibility to execute JavaScript in the page to interact with it, for example to show content of interest otherwise hidden (ref. \arch{2}, Figure~\ref{fig:train_js}, in Appendix). The training interface allows the user to specify the type of the current page (login, homepage, section, optionally subsection, and thread), which is used to learn the target structure. After indicating the correct page type, \tool\ will show the specific labels relevant for that page. By clicking on a label, it is possible to apply it to the relevant element(s).\footnote{To keep the training procedure as flexible as possible, \tool\ does not mandate the training of any specific element(\req{4}).} Once the user identified each relevant element in displayed page, they can confirm the selection. 

\smallskip
\noindent
\textit{Identifiers computation.} The identified elements are processed to produce \textit{XPaths} identifiers. \textit{XPaths} are calculated using four strategies, each working as fallback to the previous one. Each element's XPath is independently calculated with the first strategy yielding a correct result; the four strategies are:

\begin{itemize}
\item The \textit{first strategy} is an implementation of the algorithm proposed by Leotta et al.~\cite{leotta2016robula+}, that prioritizes a set of attributes stably identifying HTML elements (e.g., \texttt{id}, \texttt{name}, \texttt{class}, ...), avoiding those who do the opposite (e.g. \texttt{src}, \texttt{href}, \texttt{height}, ...). However, in the context of anti-crawler measures, IDs and names may be randomized, resulting in an apparently valid training that is no longer effective in a new session or on page reload. Also, this strategy can only calculates the identifier for a single element, while in many cases we may need to have an identifier matching several elements. 

\item The \textit{second strategy} accounts for these limitations and tries to calculate XPaths by constructing the absolute XPath (i.e., fully characterizing all the descending selectors starting from the root element \texttt{/html}). If two or more elements are provided, the strategy calculates the absolute XPaths for all of them and derives the common XPath matching them all. 

\item The \textit{third strategy} attempts to extract an element's \texttt{class} attribute in the case of randomized identifiers and full XPath changing (e.g., the number of navigation buttons changing per thread depending on their number of pages). While this can be also performed from the first strategy, it is not capable of deriving the common classes for different elements. This method therefore attempts to calculate the common \texttt{class} attribute of different elements. 

\item The \textit{fourth strategy} tries to calculate the XPath using Selenium; for the obtained identifiers from the training, Selenium finds the related \texttt{WebElement} object and extracts their XPaths. In the case of multiple elements, this strategy uses the second strategy's approach to derive a common XPath.
\end{itemize}
As a last resort, in case all the four strategies fail to deliver an XPath, \tool\ will ask to provide an XPath identifying the problematic element(s) in the displayed page. This requires the user to manually calculate a stable XPath.

\smallskip
\noindent
\textit{Verification.}
Once all the submitted elements yielded an XPath, a new confirmation window will render again the page, color-coding each relevant element for visual verification. The user can verify if the training was correct; if one of the XPath calculation strategies yielded a wrong result it is possible to adjust the learned structure, by correcting the wrongly labeled elements. This will cause the next strategy to run and calculate new XPaths. In the case of XPaths matching more elements than desired (e.g., only specific sections identified by their \texttt{tr} element, rather than all \texttt{tr} elements inside of a \texttt{table}), \tool\ offers the possibility to (de)select all elements that should not be included in the current selection. Once the page is correctly labeled, the training for the current page is considered complete; the page is then reloaded and \tool\ checks if the calculated identifiers are stable. 
If not, a prompt will ask if the user can still see the element that \tool\ could not find. If this is the case, \tool\ considers the element identifiers in the page as unstable (i.e., likely randomized) and the training continues until a strategy provides stable identifiers. When the identifiers are deemed stable, it is possible to move to the next page in the queue. For pages containing navigation items (i.e., next or previous page buttons) \tool\ will load also the next page by clicking the next page button, to verify that the right strategy was used and the training was successful; these elements are particularly sensitive as their position and number varies when moving to the next page. 
When one page contains multiple elements of the same category (e.g., thread titles), it is possible to click multiple of them; in this case, the training module employs the first strategy capturing the XPath matching all the identified elements. Finally, for elements containing dates (e.g., post date), it is possible to specify a date format for parsing. Once the training for all pages is terminated, the procedure is completed.



\paragraph{\arch{2}. JavaScript injection module} 
Some underground platforms are particularly difficult to crawl, as they dynamically load content in the page upon interaction without affecting the URL, making the desired content unreachable at that stage. In the training page, \tool\ offers the possibility to inject and execute JavaScript in the loaded page via the \texttt{execute\_script} function of \textit{tbselenium}~\cite{tbselenium}, a browser instrumentation library extending the popular browser instrumentation library \textit{Selenium}~\cite{selenium} to support TOR Browser. 
After the script execution, \tool\ proceeds to render the updated page again in the training interface. Upon confirmation of the training, the JavaScript code is saved in the database. Every time a page of the same type is loaded during the crawling, \tool\ will execute the script before interacting with it.
Among other use cases, the JavaScript injection module allows to remove elements hindering the interaction with the page (e.g., closing a popup) or to show the list of sections of interest. The module offers a button that pre-generates the needed code to click on an element, and the user needs only to identify its XPath and replace it in the code (Figure~\ref{fig:train_js}). More advanced cases include accessing to `private' forum sections, where a page requiring an additional password may be displayed, or showing hidden post content after multiple interactions with it. 

\paragraph{\arch{3}. Scheduler}
During the setup of \tool\ it is possible to define a schedule for the crawler. The scheduler allows to specify when the crawler should start and end its execution over each weekday, and to schedule pauses in between a crawling session. Also, it is possible to specify how strictly the schedule must be followed, by defining ranges that alter the start and end time of both crawler activity and pauses, and the timezone to which the schedule applies. When the crawler is allowed to start, the scheduler compiles a list of time spans for the crawler to run or pause.

\begin{figure*}
    \centering
    \includegraphics[width=0.94\textwidth,keepaspectratio]{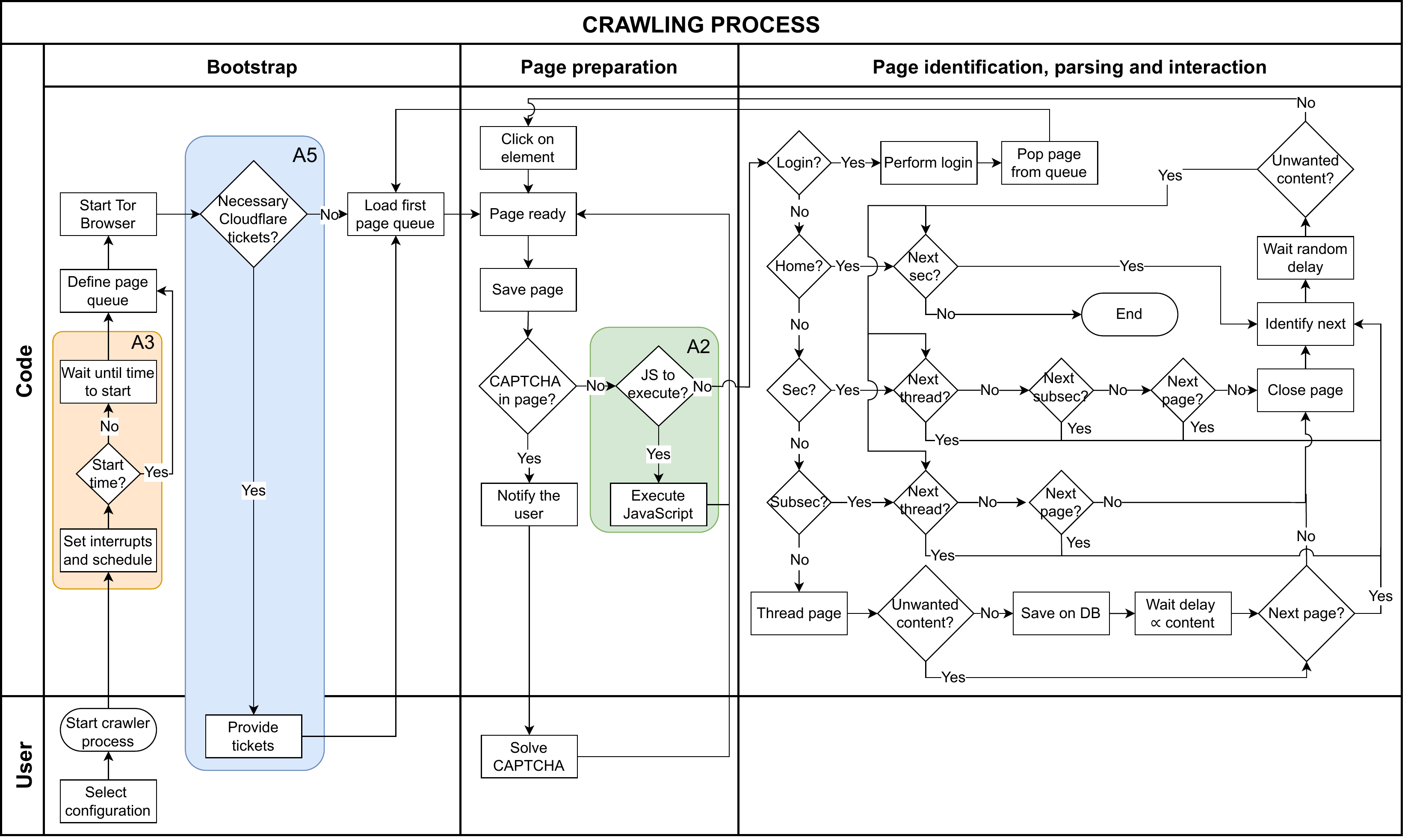}
    \caption{Crawler execution process diagram.}
    \label{fig:crawl_process}
\end{figure*}

\paragraph{\arch{4}. Crawling module} 
The crawling process is summarized in the process diagram in Figure~\ref{fig:crawl_process}. 
The crawler uses an instrumented instance of TOR Browser~\cite{torbrowser}, maneuvered via \textit{tbselenium}. 
We decided to use TOR Browser to improve the anonymity of \tool, while granting access to underground communities available over TOR. Selenium uses \texttt{geckodriver} to hook to TOR Browser's APIs; however, it discloses that the current browser is controlled by automation by setting a read-only variable \texttt{navigator.webdriver} to true. To avoid this, we create a profile for TOR Browser coming with the extension \texttt{TamperMonkey}, which allows to create and execute scripts during the lifecycle of a webpage. 

\smallskip
\noindent\textit{Bootstrap.} The crawler can start in two different ways: it can begin after completing a training procedure for a new forum, or from a pre-existing configuration of interest for which training already happened. In both cases, the GUI spawns the crawler process by passing the relevant configuration. The scheduler calculates when the crawler should start and end the activity, and schedules both the breaks defined during the configuration and a number of random interrupts. When it is time to start, it creates a queue of pages to visit (namely, the login page and the home page only) and starts TOR Browser. Similarly to \arch{1}, it asks the user if Cloudflare tickets are necessary. Then, it loads the first page in queue.

\smallskip
\noindent\textit{Page preparation.} When the (login) page is ready, the crawler saves it and checks for the presence of CAPTCHAs; if present, the user will be prompted to solve it. Then, the module checks if for that page there is some JavaScript to execute. If so, the page is saved again and it is ready to be parsed.

\smallskip
\noindent\textit{Page identification, parsing and interaction.}  In the case of a login page, login is executed and the next page loaded is the home page. When moving across pages that are not thread pages, \tool\ waits a random number of seconds between 5 and 15. For each section (and optionally subsection, if specified), \tool\ checks if there are threads yet to crawl. If so, each thread of interest (that is, not containing blacklisted keywords in the title) is opened in random order, one at a time, and traversed to extract its content, which is sent to the database. Before moving to the next page of a thread, \tool\ calculates a waiting time based on the WPM (words-per-minute) speed defined in the configuration and on the length of the text content of the current page. 
When the thread crawling is completed, the crawler returns to the parent section (or subsection) and looks for a new thread to crawl. If all threads in the page have been crawled, the crawler will attempt to reach the next page of the current (sub)section. When completed, it moves to the next one, if any. The crawler can suspend its execution for the scheduled pauses (randomly) planned during the definition of the schedule. When all sections have been fully crawled, the crawler will terminate its execution. 

\paragraph{\arch{5}. Privacy Pass ticket injection module} An increasing amount of underground communities are adopting Cloudflare DDoS protection to mitigate attacks to their infrastructure. Lately, Cloudflare CAPTCHAs have become particularly obtrusive when trying to access a protected website via TOR, due to the low reputation assigned to IPs of TOR's exit-nodes'. This causes the presence of very long sequences of CAPTCHAs. In 2018, a group of researchers developed a security-enhancing protocol and an extension in cooperation with Cloudflare that allows users to solve CAPTCHAs in exchange of so-called `tickets' that can be used to bypass Cloudflare's CAPTCHAs~\cite{davidson2018privacy}. 
\tool\ comes with a TOR Browser profile with the \textit{Privacy Pass} extension installed, that allows to legitimately earn tickets from Cloudflare's website \texttt{captcha.website}, and to store them in the extension. The only caveat is that earning tickets is not possible via TOR Browser because \texttt{captcha.website} is protected from the same DDoS protection mechanism, requiring to browse the website from the clear web. The user can provide the obtained tickets to \tool\ via the dedicated interface.\footnote{Privacy Pass functionality is suspended by Cloudflare when Cloudflare customers declare an ongoing attack (``I'm under attack!'' mode); this provides various mitigation techniques to DDoS attacks, among which disabling the Privacy Pass protocol~\cite{cloudflare}. This causes the extension not to be effective and the interstitial CAPTCHA page to be displayed.} 

\section{\tool\ evaluation against live, active underground forums}
\label{sec:evaluation}

We tested \tool\ against seven live, active criminal underground forums to which we have access. In this Section, we start by providing an overview of these forums and the overall capability of \tool\ to adapt to the different environments to crawl. We then discuss in detail the capabilities of the tool's architectural components in relation to the most interesting challenges posed by the selected forums. Finally, we provide an overall description of the performance of \tool\ across the forums. The present evaluation serves two purposes: (1) to evaluate the effectiveness and performance of the proposed core functionalities of the tool; (2) to identify weaknesses and points of improvement for future iterations of the tool, as some edge-cases not considered at design/implementation time may not be fully supported yet. This data collection was performed under ERB approval ERB2021MCS1.

\subsection{Selected underground forums}

Table~\ref{tab:forums}
\begin{table*}
\centering
\caption{Descriptive statistics for the seven selected underground forums.}
\label{tab:forums}
\scalebox{0.92}{
\small
\begin{tabular}{lllllllll}
\toprule
& \textbf{Focus} & \textbf{Language(s)} & \textbf{CAPTCHAs?} & \textbf{Sec} & \textbf{Subsec} & \textbf{Threads}  & \textbf{Posts} & \textbf{First activity} \\                             
  \midrule
\crdclub & Carding, documents, fraud & EN, RU & No & $4$ & $47$ & $86'537$ & $395'276$ & Jul 8, 2016$^\dagger$ \\
\nulled & Leaks, accounts, fraud & EN & Yes & $48$ & $45$ & $1'203'886$ & $35'177'498$ & Apr 22, 2015$^\dagger$ \\
\xss & Malware, spam & RU & No & $48$ & $3$ &$50'610$ & $394'486$ & Sep 19, 2018$^\ddagger$ \\
\altenen & Ewhoring, malware, accounts & EN & Yes & $55$ & $68$ & $970'023$ & $6'840'943$ & Mar 22, 2010$^\dagger$ \\
\nulledbb & Accounts, hosting, Ewhoring & EN & No & $23$ & $61$ & $\sim206K$ & $\sim1.5M$ & Jan 01, 2015$^\dagger$ \\
\deeptor & Carding, fraud & EN & No & $31$ & $8$ & $14'132$ & $98'631$ & Jul 24, 2015 $^\dagger$ \\
\darknetcity & Accounts, proxy, fraud & EN & No & $53$ & $0$ & $3'089$ & $15'018$ & Oct 26, 2017$^\dagger$ \\
\bottomrule
\multicolumn{8}{l}{\small{Information fetched on June 18th, 2022. $^\dagger$: oldest staff registration date. $^\ddagger$: domain registration date.}} 
\end{tabular}}
\end{table*}
\begin{table*}
\centering
\small
\caption{Summary of \tool\ performances across the selected forums.}
\label{tab:performance}
\scalebox{0.94}{
\begin{tabular}{lcccccccccc}
\toprule
& Train & Crawl & Time\textsubscript{total} & Time\textsubscript{breaks} & WPM & Download img & JS exec & CF tickets & Threads & Posts \\
\midrule
\crdclub\ & \cmark & \cmark & 4:05:52 & 40:45 & $180-240$ & \cmark & \xmark & \xmark & $1$ & $330$ \\
\nulled\ & \cmark & \xmark & - & - & $180-240$ & \cmark & \cmark & \cmark & - & - \\
\xss\ & \cmark & \cmark & 3:51:40 & 42:46 & $180-240$ & \cmark & \xmark & \xmark & $1$ & $580$ \\
\altenen\ & \cmark & \cmark$^\dagger$ & 3:31:12 & 28:29 & $180-240$ & \xmark & \cmark & \xmark & $94$ & $1'691$ \\ 
\nulledbb\ & $\sim$ \cmark & \xmark & 1:04:54 & 00:00 & $180-240$ & \cmark & \xmark & \xmark & $4$ & $13$ \\
\deeptor\ & $\sim$ \cmark & \xmark$^\ddagger$ & 08:12$^\ddagger$ & 00:00 & $180-240$ & \cmark & \xmark & \xmark & $1$ & $10$ \\
\darknetcity\ & \cmark & \cmark$^\dagger$ & 3:31:29 & 44:15 & $600-800$ & \xmark & \xmark & \xmark & $6$ & $1'451$  \\
\bottomrule
\multicolumn{11}{l}{\small{$^\dagger$: premature termination due to connectivity issues with the target; $^\ddagger$: manual termination of the tool due to wrong behavior during crawling.}}
\end{tabular}}
\begin{minipage}{0.95\textwidth}\footnotesize
\smallskip
All the parameters specifying the page loading and download duration timeouts, the timezone adopted, the variance intervals to apply when calculating the start and end of workday and breaks, as well as interruption duration and minimum time between two interruptions are set to default. 
\end{minipage}
\end{table*}
provides an overview of the selected forums for the evaluation. All forums have been active for at least four years, with the oldest recorded activity in Jan 2015 for \nulledbb. Almost all are organized in sections and subsections, although numbers vary widely across forums as do the number of posts. Selected forums are also well varied in terms of number of posts and cover English and Russian locales. \nulled\ and \altenen\ employ a CAPTCHA system at login time. Below we discuss the forums' relation to the problem space defined in Section~\ref{sec:probspace}.
\crdclub\ provides a baseline for the performance evaluation. This forum does not come with any specific anti-crawler measure and its structure is rather straightforward. However, it features two inconvenient aspects: it shows a popup when a direct message is received, altering the interaction with the page, and it opens the last page of thread if we click on its title. The former problem is solved by disabling this option in the user control panel of the website, while the latter, shared with \nulledbb, is tackled with a specific solution implemented in \arch{1}. 
\nulled\ represents the benchmark to test both \arch{2} and \arch{5} capabilities; respectively, marketplace sections can be dynamically loaded after clicking on a button in the homepage, and the website is protected by Cloudflare DDoS protection. \xss\ implements ID randomization in the DOM and some elements such as thread title in the subsections and post author details do not come in predictable positions all the times, which is interesting from \arch{1} perspective. \altenen\ is a forum requiring to interact multiple times in a thread to show the hidden content (\arch{2}), causing threads to be long and rich in spam, offering interesting considerations during the execution of \arch{1}. 
\nulledbb\ presents an interesting marketplace section, and we configured \tool\ to target content presenting malware related terminology. 
Finally, \darknetcity\ is a forum hosted on TOR, and comes with a non-trivial layout for the user details in a post. In addition, \altenen\ and \darknetcity\ present significant performance issues due to the large amount of high-definition GIFs, worsened by TOR's bandwidth; we present a solution to mitigate this issue. 
We also discuss the problems encountered during the training (\nulledbb) and the crawling (\deeptor, \nulledbb\ and \nulled). 

\subsection{Overall performance}


We test \tool\ functionalities by performing training and crawling, for a session of four hours, for all forums. Rather than completing the data collection, our goal here for the presented prototype is to test whether the core functionalities of \tool\ work across all forums and are sufficient, to identify those who need further refinement and to discuss about the possible alternatives, while providing an estimate figure of the volume of crawled pages in a unit of time. To provide a range of estimations for different use cases, we customize configurations for each forum depending on the expected quantity of content, long delays from the platform, and desired stealth. Table~\ref{tab:performance} details the chosen configurations and summarizes \tool\ performance. The relatively low number of threads and posts visited for \crdclub\ and \xss\ compared to \altenen\ and \darknetcity\ can be traced back to the verbosity of their posts. Often, members quote the original post of the author, increasing the delay before moving to the next page. Considering that a regular user would notice the repeated content, an higher WPM range could be defined, as we did for \darknetcity. Both \altenen\ and \darknetcity\ suffered from a premature termination of the crawling session due to connectivity issues with the websites. However, the results were interesting; both performed well, collecting $1'691$ and $1'451$ posts across $94$ and $6$ threads respectively in approximately three hours and a half. As mentioned, both forums feature a large amount of high-definition GIFs, making the complete loading of the page extremely long and causing timeouts. In the configuration, \tool\ allows to disregard loading images and, albeit representing a possible suspicious behavior, the solution proved effective. 
\altenen's training had a visual glitch, where clicking on the ``next page'' button would highlight it temporarily, but the training was not negatively affected. For this forum we also created some JavaScript to show hidden content in the threads by liking and replying to the original post. Finally, we could not perform an adequate crawling session on \nulled. Despite the presence of valid Cloudflare tickets, the forum set the ``I'm under attack!'' mode, disabling the functionalities of the extension~\cite{cloudflare} and making crawling impossible for more than a few minutes. The training for \nulledbb\ was complicated due to an improper rendering of the identified elements, making the adjustment process tedious until the problematic label was identified, a stable identifier was manually created and provided via the fifth strategy.
Whereas \deeptor\ successfully completed the training session, its structure seemed to change during crawling, thus making the tool incapable of accessing previously visited threads.

Overall, \tool\ was successfully trained for all the forums, with some imperfections for two of them, and managed to crawl four of the seven live forums employed for the evaluation. A fifth one, \nulled\, could be theoretically crawled, but the defences in place at the moment of the benchmark blocked TOR IP addresses, obstructing our operations. The problems encountered in the two remaining forums will be discussed in relation to the appropriate architectural component in Section~\ref{subsec:tech_eval}.

\subsection{Technical rundown}
\label{subsec:tech_eval}
In this Section we provide an insight of the involved processes from the tool's perspective, and detailing the most interesting cases. 
\subsubsection{\arch{1} Training} 
The underground forum \xss\ implements randomized IDs for several elements across its pages as a anti-crawler countermeasure. In the login page, the username and password fields have \texttt{id=\_xfUid-1-\textbf{timestamp}}, where \texttt{timestamp} is expressed in seconds from epoch. In this case, these elements also present a stable attribute \texttt{autocomplete}, which is identified from the first strategy and used as a reliable identifier. In the home page, 
the user proceeds to label the sections and subsections of interest. In this case, we are only interested in the subsections ``Malware'' (\texttt{XPath: /html/.../\textbf{div[5]}/.../h3[1]/a[1]
} and ``Cracking'' (\texttt{XPath: /html/.../\textbf{div[6]}/.../h3[1]/a[1]
} under the section ``Underground''. When two or more elements of the same type are defined
, \tool\ attempts to infer a common XPath that matches all the selected elements of that type using the second strategy. This is beneficial from a user perspective, as it allows to identify only a few examples to infer the identifier, instead of clicking them all (e.g., the list of all threads in a page). However, in that case we are interested in only that specific set of elements; it is possible to click on the button ``Ignore'' and select the elements to blacklist (i.e., the uninteresting subsections). The strategy in charge (strategy two) obtains the correct XPath (\texttt{XPath: /html/.../\textbf{div}/.../h3[1]/a[1]} while keeping a list of the ignored XPaths, and thus telling \tool\ to access only those matching the XPath that are not blacklisted.
\sloppy In subsections, threads can be generally identified by using the XPath \texttt{/html/.../div[\textbf{thread\_id}]/div[2]/div[1]/a[1]}. However, sometimes they present a tag before their name, resulting in tagged threads being identified by the XPath selector \texttt{/html/.../div[thread\_id]/div[2]/div[1]/\textbf{a[2]}}. Selecting both types of threads would generate the common XPath \texttt{/html/.../\textbf{div}/div[2]/div[1]/\textbf{a}} with strategy two (note that the div in bold matches all the thread\_ids), which matches both the thread links and tag links. 
Once we acknowledge that every subsequent strategy fails to identify the list of threads correctly, we are prompted to provide an XPath that we can calculate by inspecting the page (\req{4}). From manual inspection it is possible to note that thread titles consistently present the attribute \texttt{data-xf-init="preview-tooltip"}. It is possible to use this attribute to generate the XPath \texttt{//*[@data-xf-init="preview-tooltip"]} and to provide it to the trainer, thus solving the problem.


The training of \nulledbb\ resulted challenging in the thread page. One or more wrong identifiers matched wide areas of the page; for example, the post date identifier was wrong and resulted in a verification window without any labeled element. This made the training complicated, as it was not possible to correct the labeling for the single wrong identifier, leading the tool to use new strategies even for the correct identifiers. After some attempts, it became clear that also the identifiers for the post content were creating problems during the rendering of the identified elements. After reiterating the training for a few times, \tool\ asked to provide the XPaths for the problematic elements. From manual inspection of the page, we identified the attribute \texttt{data-original-title='Original post time'} for the post date, and \texttt{class='post-message flex-fill} for the post content, and we created the corresponding stable identifiers. 

\subsubsection{\arch{2} JavaScript injection module}
\label{subsec:casestudy-js}
\altenen\ allows to see the content of a post after the user ``likes'' the post and replies to it as an anti-crawler measure. To solve this problem, during the training of a thread page, the user can write a script to perform these actions. The like, quote, and send reply buttons come in predictable places, \texttt{//post\_footer/div[1]/div/a[1]/span/bdi}, \texttt{//p\_footer/div/div/a[2]}, and \texttt{//form/../button[1]/span}
respectively.
The user needs to prefill the JavaScript injection box with the ``click on element'' code (ref. Figure~\ref{fig:train_js}) and to provide the relevant XPaths. Considering the time required to submit the like, quoting the post and sending the reply, it is necessary to introduce some waiting time between actions, by using the \texttt{await new Promise(r => setTimeout(r, \textit{millis}))} function. 
Once ready, the script is executed in the page in TOR Browser via \textit{tbselenium}, and the page is downloaded and rendered again. Upon confirmation of the training, the JavaScript code is saved in the database.

\nulled\ organizes the content in several sections accessible from the home page by clicking to the ``topic'' of interest, which shows the relevant sections. This content cannot be accessed directly from a URL, but rather requires the user to click on the topic of interest. Similarly as seen in \altenen\, to click on the ``Leaks'' topic button, it is sufficient to prefill the JavaScript injection box and provide the button's XPath. 

\subsubsection{\arch{3} Scheduler}
\label{subsec:casestudy-scheduler}
The scheduler execution does not have noteworthy details to report for the evaluation set.


\subsubsection{\arch{4} Crawler}
\label{subsec:casestudy-crawler}




\altenen\ contains unwanted material, such as so-called revenge pornography material. We set the crawler to explore all links but to avoid threads containing the keywords ``GF'', ``nudes'', ``photos'', ``snapchat'' and ``naked'' in any of their posts or in the title (Figure~\ref{fig:keywords}). To achieve that, the crawler parses the current page (i.e., a section), seeking for threads to explore. From the threads list, it checks if any of these should be excluded based on the blacklisted keywords. The same process applies while browsing a thread: if any of the posts mentions any of the keywords, the thread is closed, the posts discarded, and the tool moves to the next thread. Similarly, \nulledbb\ features potentially interesting content among a large amount of spam, and we set the crawler to explore only threads matching one or more relevant keywords related to malware trade. However, crawling for \nulledbb\ failed due to Selenium being unable to detect if the browsed page was successfully loaded or not. This resulted in a page not successfully loaded and Selenium not raising a timeout error to let \tool\ reload the page and try again, ultimately stalling the crawler.  

\deeptor\ crawling failed in the section page. The problem is that the position in page for thread titles mutates when a thread is accessed for the first time; when \tool\ returns to the section page after crawling the first thread, it fails to identify all the threads in the page, resulting in an error and prematurely terminating the crawling of the current section. A solution could be to manually inspect the structure of the page to derive XPaths matching both cases. XPath syntax includes an UNION operator, which could be used to derive the list of threads for both cases. Therefore, running again the training and voluntarily falling back to the XPath injection strategy (ignoring the currently correct training not accounting for the future DOM of the page) is a possible workaround. 
\nulledbb\ crawling stalled when a page failed to load; this is not an uncommon issue within the crawling context, especially when using TOR, and \tool\ manages this issue by interpreting the errors arising from Selenium. However, during this run we encountered a case in which Selenium ``hangs'' indefinitely, and our tool manages the situation as a network issue, and attempts to refresh the page. Despite that, Selenium remains unresponsive, and our tool cannot proceed in the crawling. This issue would require to inspect Selenium and to extend its functionalities. 

\subsubsection{\arch{5} Privacy Pass ticket injection module}
\label{subsec:casestudy-privacypass}
To both train and crawl \nulled, we need to prevent Cloudflare to show the CAPTCHA page. To do so, the user has first to earn tickets on \texttt{captcha.website} in the case of Cloudflare and then to export them. The user needs to access to the Firefox debug mode (\texttt{about:debugging\#/runtime/this-firefox}) and click on inspect for Privacy Pass. By browsing the `Storage' tab, under `Local Storage' they can find the tickets in the form of two key-value pairs (\texttt{cf-commitment-2.58} and \texttt{cf-tokens}) to copy and paste into \tool's dialog, (Figure~\ref{fig:tickets}).
After submitting all the key-value pairs, \tool\ will trigger a sequence of actions via Selenium, opening the same page and executing JavaScript in the browser console, thus loading the tickets in the extension. Once this operation is completed, the browser will successfully load the target website, bypassing the CAPTCHA. 
This procedure allowed us to perform the training of \nulled; however when we tried to crawl some days later, the forum set the ``I'm under attack!'' mode and stopped accepting tickets (considering the timing of the episode, and the negligible volume of traffic generated by our training we consider it unlikely that our training session caused the state change). A possible solution for a future release of \tool\ would be to allow the option of using either TOR Browser or Firefox for the execution of \tool\ with a dedicated proxy, thus avoiding to use the IP addresses of TOR exit-nodes which generally suffer of bad reputation.

\subsection{User interface}
\label{sec:gui}

\subsubsection{\arch{1}. Training module}
In the \textit{(Bootstrap}) phase of this module, the GUI 
offers the user the possibility of creating a new configuration for a new forum or to use an existing one. The configuration interface is shown in Figure~\ref{fig:config_interface}. 
When the crawler is allowed to start, it asks for Cloudflare tickets and finally loads the login page.
Additionally, the ``Keywords'' tab allows the user to specify keywords indicating unwanted content or relevant keywords, allowing the crawler to respectively avoid opening threads containing any of these in the title or in any of its posts, or to open only those containing them in the title in the case of relevant keywords (Figure~\ref{fig:keywords}). 
The user can now proceed to label the page, selecting the relevant label and clicking on the corresponding element. 

Upon submission, a new window will appear to confirm the current selection, showing the identified elements obtained from \tool\ with the learned information. 
The user has the chance to manually remove the wrong labels or to reset completely the training to start from scratch, and to add them once again. 
The user can reiterate the process until the result is satisfactory. As a last resort, \tool\ may ask the user to provide an XPath 
to identify the specified element(s). 
Once an XPath is provided, \tool\ renders again the page to confirm the selection. 

\subsubsection{\arch{2}. JavaScript injection module}
When a script was defined during the training for the specific page, it is executed before interacting with the page. Figure~\ref{fig:train_js} in the appendix shows the interface to inject the desired JavaScript as discussed in Section~\ref{subsec:casestudy-js}. The currently displayed code is generated from the ``prefill'' button, which allows to identify an element in the page via XPath and to click on it; the user is then left to replace ``YOUR\_XPATH\_HERE'' with the correct XPath. 

\subsubsection{\arch{3}. Scheduler}

The schedule for a crawling activity is defined at configuration time. Figure~\ref{fig:scheduler} in the appendix provides a view of the interface. 
The indicated times apply to the specified timezone defined in the configuration tab. In our case the crawler should perform its activity during weekdays from 17:00 to 20:00; during weekends, the crawler should work from 9:30 to 13:30 with a scheduled pause between 10:30 and 11:00 (this is the schedule used for the benchmark), and another crawling session from 15:00 to 20:00. 

\subsubsection{\arch{4}. Crawling module}
During the crawling process, one of the two interactions with the user is the notification of a CAPTCHA in the page. During the login on \nulled\ and \altenen\, the crawler informs the user that there is a Google reCAPTCHA in the login page and asks to solve it. When the CAPTCHA is solved, the user confirms by typing ``solved'' in the terminal and the crawler proceeds to login. 
The second interaction available to the user is the manipulation the execution of \tool\ by typing in the console the commands ``resume'' to skip the current break, interruption or delay, resuming the crawling immediately, ``pause'' to suspend its execution or ``terminate'' to end the current crawling session.

\subsubsection{\arch{5}. Privacy Pass ticket injection module}
\nulled\ is protected by Cloudflare DDoS protection. Therefore, accessing it via TOR Browser will cause Cloudflare to show the CAPTCHA page and deny access due to TOR's exit-nodes poor reputation. To solve this issue, at startup the crawler asks whether the user wants to provide Cloudflare tickets before execution. To earn tickets, it is sufficient to open a regular Firefox instance with the Privacy Pass extension installed and browse the website \texttt{captcha.website}. Solving a challenge will grant the extension 30 tickets necessary to bypass the CAPTCHA page (which could be displayed several times during one session) and consume more than a single ticket to bypass it. Solving the challenge multiple times allows to earn more tickets, granting access for a longer session. These tickets can be exported and provided in the prompt of \tool, as shown in Figure~\ref{fig:tickets}.

\section{Discussion and conclusion}
\label{sec:discuss}

In this paper we presented \tool, a general method to learn and stealthily crawl arbitrary (underground) forums. We present the foundational challenges, proper of the problem space, such a solution must address, and design our method and overall solution around those. We provide a prototype implementation and test it against live, underground forums.
The results show that \tool\ successfully managed to learn
all the forum structures and to crawl four out of seven forums 
proposed for validation. The tool successfully learned the structure and content layout of the forums using different strategies accounting for possible anti-crawler measures. \nulled, \xss\, \darknetcity\ and \nulledbb\ showcase one of a module allowing the user to provide a manually created identifier when all the identification strategies of the tool fail. The crawling sessions, configured to last four hours terminated successfully in four cases. Out of these four, \tool\ was configured to not download pictures from two targets, \darknetcity\ and \altenen, mitigating the long loading times during the crawling and reducing TOR network stress; in addition, we modified the WPM parameter to shorten the delay between the crawling of two pages, considering the large amounts of spam in the posts of \darknetcity. \darknetcity, \nulledbb\ and \deeptor\ do not feature any subsections, and the crawler manages this case naturally. \crdclub\ and \nulledbb\ present a special case in which clicking on a thread opens its last page; the tool offers the possibility to handle this case by training the button opening the first page of the thread when landing on it. In another two forums, \nulled\ and \altenen, we showcased another feature supporting the extensibility of the tool, the JavaScript injection module; for the former, we wrote a script to reach the sections of interest from the front page, while for the latter we wrote a script to interact with the page to reveal the hidden content of a post. Furthermore, during the training of \altenen\ we opted to not to train the next page button, causing \tool\ to seamlessly terminate the crawling of the current thread and move to the next one, to avoid crawling pages containing spam. Moreover, \nulled imposed a strict policy for visitors, nullifying our efforts of bypassing Cloudflare's CAPTCHAs during the benchmark. However, this is a problem affecting every user of the platform that tries to access it via TOR. A possible solution to this problem would be to offer \tool\ the possibility of switching to a regular Firefox browser, using an appropriate proxy for anonymity and benefiting from a `clean' IP address. \deeptor\ had an apparently correct training, but during the crawling the tool could not find the location of threads within the section; this was caused from a change in the DOM after the thread was opened. However, the problem could be solved by rejecting the apparently correct training for the problematic element until the XPath injection module is triggered, and providing an XPath accounting for both cases. Finally, \nulledbb\ had problems in rendering the calculated identifiers during the training due to one or more wrong identifiers; this could be solved by implementing a feature that allows to visually verify identifiers one at a time, narrowing down the retraining to the incorrect identifiers only and avoiding to run the training several times to guess what are the problematic labels. On the same platform, we experienced a limitation of the Selenium framework, which turned to be incapable of detecting whether a page was successfully loaded or not, and stalling the execution of \tool. 


\subsection{\tool\ release}
A significant portion of today's research on cybercriminal communities relies on leaked data and old datasets, allowing to perform \textit{post-mortem} analysis on them. In other cases, researchers develop \textit{ad-hoc} crawlers and parsers to tap data from each community of interest, which is a burdensome procedure. This software is rarely shared among the community, because its purpose is limited to the scope of the research. The cost of developing such software discourages research and limits its scope, whereas the (in)success of extracting data from a community can make the whole research unfeasible. Risks are higher when the target is a prominent community, where the costs (e.g., pecuniary) of losing access are substantial and obtaining access again may be ethically hard to justify. 

Therefore, we propose and release \tool, a prototype crawler that aims to address different problems related to the crawling of criminal communities, offering a supervised procedure to learn the structure of a target community, supporting manual intervention in particular cases, and enabling dynamic interaction with the page via JavaScript to circumvent several custom anti-crawler mechanisms. Together with a (potentially more supported) CAPTCHA bypass mechanism and the modelling of a seemingly-legitimate user, \tool\ proves that it is possible to crawl across a number of different underground communities, without the burden of creating scarcely reusable software for both crawling and parsing their content while remaining stealth, and giving the user possibility to tune the tool to reach the desired trade-off between stealth and throughput.

\subsection{Future work}
\label{subsec:future}

As the tool is currently a prototype meant to showcase the overall approach and its viability, from the evaluation we identify a number of key improvement points to address in the future, also for possible uptakes from the community.


\paragraph{Training procedure} The training of \nulledbb\ was particularly complicated due to the improper rendering of the calculated identifiers. 
The training interface could be improved to ease the troubleshooting of these problematic training scenarios, allowing to highlight one family of identified elements at a time. Detecting only CAPTCHAs hinders the crawling of a target platform. A fully-fledged solution would require to detect and forward CAPTCHAs to an operator in charge of solving them to resume the crawling of the target platform. Commercial solutions managing CAPTCHA resolution exist, and could be used for the same purpose. At its current state, the crawler does not support the crawling of threads that have been already visited. This could be solved by training in the section/subsection \tool\ to recognize the count of posts within a thread and to keep track of it; if this number does not correspond to the stored value during the previous crawling session, the thread should be visited again. This could be an optional feature working only when this information has been provided to \tool\ during the training.

\paragraph{Crawler robustness} In the case of \nulled, albeit \arch{5} offers a solution preventing CAPTCHAs being displayed and Privacy Pass being potentially adopted by different DDoS protection services in the near future, it falls short when stronger access policies are enabled, preventing access to users connecting via TOR. To solve that, \tool\ should offer the possibility to use a regular Firefox instance tunneling traffic through a proxy different from TOR, benefiting from a non-blacklisted IP address. Another encountered problem is connected to the use of Selenium as the browser instrumentation library of choice. Selenium is designed to be a web application testing framework, and some edge-cases regarding connectivity issues that may arise are not taken in account in the current state of the library. Therefore, it is necessary to extend the library to allow its usage as a browser instrumentation framework for crawling over unreliable networks. 

\section{Conclusion}
\label{sec:conclusions}
In this paper, we showcased the advantages and limitations of the current iteration of \tool. This prototype shows the potential to achieve the identified goals for a reusable and extensible automated crawler for underground communities, highlighting the strengths of an extensible training process tackling different anti-crawler measures. We believe that addressing the recommendations described in Section~\ref{subsec:future} could make \tool\ a solid solution, capable of stealthily crawling and extracting content from a variety of underground forums, supporting research relying on datasets originating from cybercriminal communities.
In the near future, we plan to deploy an enhanced version \tool\ in our institution to start a longitudinal data collection across different criminal communities for further research in the current and active underground threat scenario. 

\smallskip
\noindent \textit{Publication, Development, and Licensing.} The development of \tool\ was partially supported by a team of BSc students as part of their final graduation project. \tool\ is released under GNU Affero General Public License v3.0. Source code and documentation are available at \url{https://gitlab.tue.nl/threat-crawl/THREATcrawl}

\section*{Acknowledgments}

This work is supported by the ITEA3 programme through the DEFRAUDIfy project funded by Rijksdienst voor Ondernemend Nederland, Grant No. ITEA191010, and by the INTERSCT project,
Grant No. NWA.1162.18.301, funded by Netherlands Organisation for Scientific Research (NWO).

\bibliographystyle{IEEEtran}
\bibliography{acmart}
\appendix

\begin{figure}[!htbp]
    \centering
    \includegraphics[width=\columnwidth,keepaspectratio]{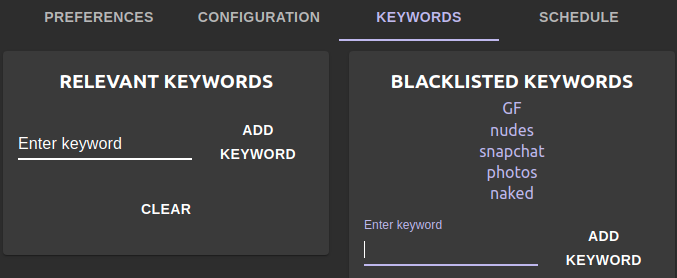}
    \caption{Keywords definition in the configuration interface.}
    \label{fig:keywords}
\end{figure}
\begin{figure}[!htbp]
    \centering
    \includegraphics[width=\columnwidth,keepaspectratio]{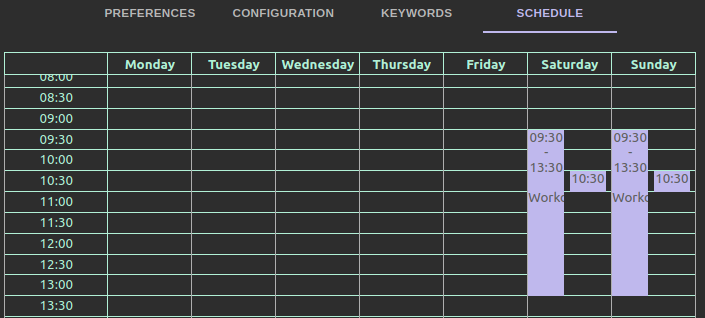}
    \caption{Overview of the scheduler configuration interface.}
    \label{fig:scheduler}
\end{figure}
\begin{figure}[!htbp]
    \centering
    \includegraphics[width=0.7\columnwidth,keepaspectratio]{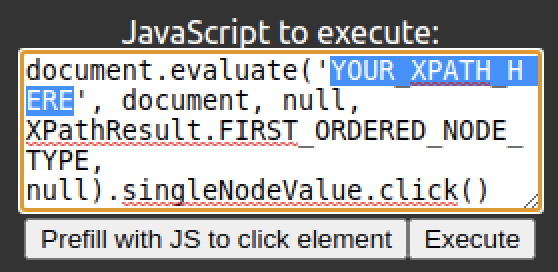}
    \caption{JavaScript injection module prefilled with a ``click-on-element'' script.}
    \label{fig:train_js}
\end{figure}
\begin{figure}[!htbp]
    \centering
    \includegraphics[width=\columnwidth,keepaspectratio]{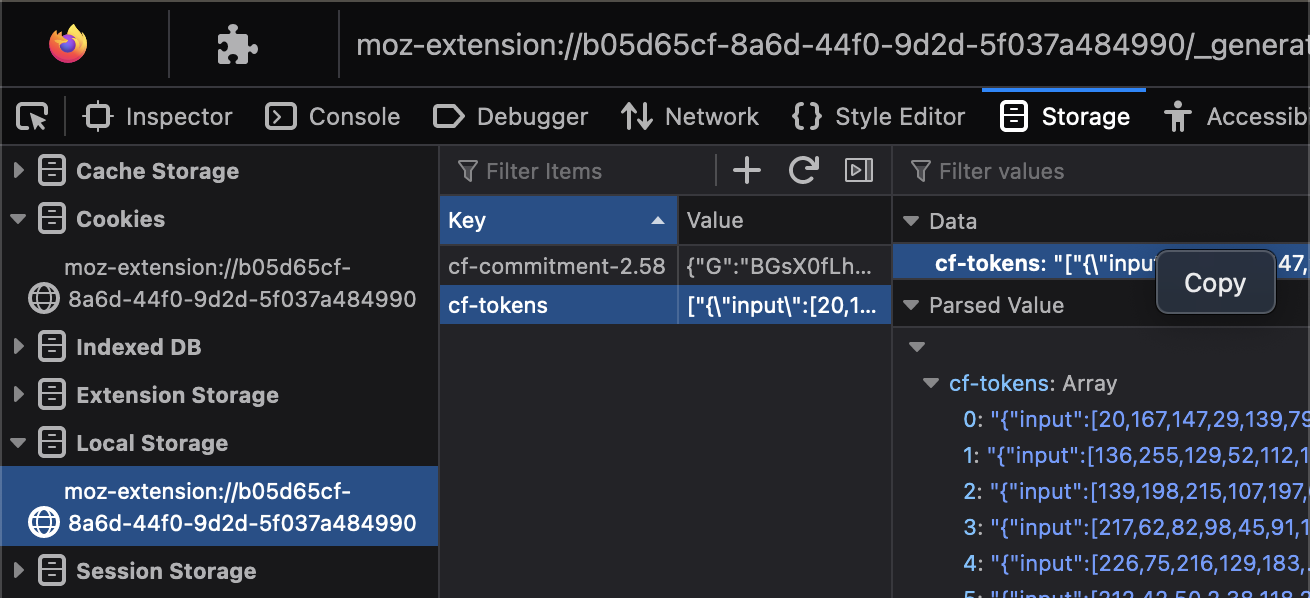}
    \caption{Screenshot of the tickets available in the source browser.}
    \label{fig:tickets}
\end{figure}

\begin{table*}
    \caption{Summary of issues and strategies addressing each identified requirement in \tool.}
    \label{tab:summaryRx}

    \centering
    \small
    \scalebox{0.9}{
    \begin{tabular}{p{0.12\textwidth} p{0.32\textwidth} p{0.5\textwidth}}
    \toprule
    Requirement & Issues & Strategy\\
    \midrule
    \req{1}. Stealth     & Crawler traffic easily identifiable compared to ``user'' behavior, (absence of) specific information in HTTP requests, or by fingerprinting the device~\cite{turk2020tight, decary2015sifting}. & Use a real browser for the crawling;
    mimic human behaviour adopting the strategy used in~\cite{campobasso2019caronte,pastrana2018crimebb} to interact with buttons and links in the displayed page, and regulate timing according to the amount of text in the page~\cite{campobasso2019caronte}. \\
    \req{2}. Configurable     &  Crawling must be limited in time during the day and should not show consistent patterns over time; not all content is relevant to crawl. & Allow user to define a crawling schedule for each day of the week, as well as when to pause. To limit the onset of patters, a number of randomly generated pauses are integrated in the schedule, and random noise added to the start and end time. 
    Specify the ``reading-speed'' of the simulated user to calculate the time between accesses to subsequent resources. 
    Define (white- and black-listed) keywords to select (or avoid) specific content. See also \texttt{MR2}.\\
    \req{3}. Trainable & Flexible identification of relevant HTML elements in a page. & Identify strategies to account for diverse forum structures and provide different solutions to infer the necessary identifiers, while offering a guided and simplified procedure to its user. \\ 
    \req{4}. Extensible  & Underground communities increasingly feature modern CMSes, supporting dynamic content generation and requiring JavaScript, making the localization of content hard to predict.  & Allow the user to input pre-generated identifiers to find HTML elements identifiers. In some circumstances, content in a page may be overshadowed or can be dynamically generated, requiring to interact first with the page; for this reason, \tool\ provides the option to perform preliminary operations by executing JavaScript on the page.\\
    \req{5}. Structured data collection     & Crawled web pages need to be parsed to extract and structure their content in an underlying database. & Allow user to label specific content to be saved from each page in a structured way. 
    \\
    \texttt{MR1}. Parsimonious     & Limit use of shared bandwidth in TOR while keeping crawling functionalities sufficiently fast. & Limit content to be crawled to focus on what necessary (R2) and throttling traffic (R1).   \\
    \texttt{MR2}. Censoring    & Avoid the download of unwanted material. & Keyword white- and black-list matching (R2).  \\
    \bottomrule
    \end{tabular}}
\end{table*}

\begin{table*}
\caption{Summary of the strategies used to derive identifiers during the training.}
\label{tab:strategies}

\centering
\small
\scalebox{0.73}{
\begin{tabularx}{1.38\textwidth}{rccccccccccccccccccc}
\toprule
& \multicolumn{4}{c}{Login page} & \multicolumn{3}{c}{Home page} & \multicolumn{6}{c}{Section page} & \multicolumn{6}{c}{Subsection page} \\
\cmidrule(r){2-5} \cmidrule(r){6-8} \cmidrule(r){9-14} \cmidrule{15-20} 
& Home & User & Pass & Login & Home & Sec(s) & Subsec(s) & Home & Sec & Subsec(s) & Threads & Next & Prev & Home & Sec & Subsec & Threads & Next & Prev \\
\crdclub\ & - & S1 & S1 & S1 & S1 & S2 & S2 & S1 & S2 & S2 & - & - & - & S1 & - & S2 & S2 & - & - \\
\nulled\ & S1 & - & - & S1 & S1 & S2 & S2 & S1 & S2 & S2 & S2 & S1 & S1 & S1 & S2 & S2 & INJ & S1 & S1 \\
\xss\ & S1 & S1 & S1 & S1 & S1 & S2 & S2 & S1 & S2 & S2 & - & - & - & S1 & S2 & S2 & INJ & S1 & S1 \\
\altenen\ & S1 & S1 & S1 & S1 & S1 & S2 & S2 & S1 & S2 & S2 & S2 & S1 & S1 & S1 & S2 & S2 & S2 & S1 & S1 \\
\nulledbb\ & - & S1 & S1 & S1 & S1 & S2 & - & S1 & S2 & - & S2 & INJ & INJ & - & - & - & - & - & - \\
\deeptor\ & S1 & S1 & S1 & S1 & S1 & - & S2 & S1 & S2 & - & INJ & S1 & S1 & - & - & - & - & - & - \\
\darknetcity\ & S1 & S1 & S1 & S1 & S1 & S2 & S2 & S1 & S2 & - & INJ & S1 & S1 & - & - & - & - & - & - \\
\midrule
\end{tabularx}}
\bigskip
\small
\scalebox{0.73}{
\begin{tabularx}{1.305\textwidth}{rcccccccccccc}
& \multicolumn{12}{c}{Thread page} \\
\cmidrule(r){2-13} 
& Home & Next & Prev & First page & Thread title & Thread sec & Post author (PA) & PA \# posts & PA popul & PA registration date & Post date & Post content \\
\crdclub\ & S1 & S1 & S1 & S1 & S2 & S2 & S2 & S2 & S2 & S2 & S2 & S2 \\
\nulled\ & S1 & S1 & S1 & - & S2 & S2 & S2 & S2 & S2 & S2 & S2 & S2 \\
\xss\ & S1 & S1 & S1 & - & S2 & S2 & S2 & S2 & INJ & S2 & S2 & S2 \\
\altenen\ & S1 & - & - & - & S2 & S2 & S2 & S2 & S2 & S2 & S2 & S2 \\ 
\nulledbb\ & S1 & S1 & S1 & S1 & S2 & S2 & S2 & S2 & S2 & - & INJ & INJ \\
\deeptor\ & S1 & S1 & S1 & - & S2 & S2 & S2 & S2 & S2 & S2 & S2 & S2 \\
\darknetcity\ & S1 & - & - & - & S2 & S2 & S2 & INJ & S2 & INJ & S2 & S2 \\
\bottomrule
\end{tabularx}}
\begin{minipage}{0.95\textwidth}\footnotesize
It is possible to see that nor S3 or S4 ever appear to be used. However, S3 exists because of \crdclub; at the beginning of 2022, we could not identify navigational items consistently with the first two strategies, and we developed this strategy to solve this issue. S4 instead has been tested in a synthetic environment, but it did not yield any useful result in the real world as of yet. 
\end{minipage}

\end{table*}

\end{document}